\documentclass[preprint,amsmath,amssymb,aps,showkeys,showpacs]{revtex4}
\usepackage[english]{babel}
\usepackage{graphicx}
\usepackage{graphics}
\usepackage{amsmath}
\usepackage{dcolumn}
\usepackage{amssymb}
\usepackage{bm}

\begin{document}
\title{A semiclassical treatment of the interaction of non-uniform electric fields with the electric quadrupole moment of a neutral particle}
\author{K. Bakke}
\email{kbakke@fisica.ufpb.br}
\affiliation{Departamento de F\'isica, Universidade Federal da Para\'iba, Caixa Postal 5008, 58051-900, Jo\~ao Pessoa, PB, Brazil.}

\begin{abstract}

By applying the WKB (Wentzel, Kramers, Brillouin)  approximation, we analyse the interaction of the electric quadrupole moment of a neutral particle with radial electric fields produced by a non-uniform electric charge densities. Then, we show particular cases of this interaction where a discrete spectrum of energy can be achieved for the $s$ waves of the neutral particle.

\end{abstract}


\maketitle

\section{Introduction}

The seminal work of Aharonov and Casher \cite{ac} has brought a great interest in quantum effects associated with neutral particles. Since then, geometric quantum phases have been investigated for neutral particles with permanent electric dipole moment \cite{hmw,hmw1}, induced electric dipole moment \cite{whw,whw2}, electric quadrupole moment \cite{chen,b} and magnetic quadrupole moment \cite{chen,fb7}. This interest has gone further in search of analogues of the Landau quantization \cite{er,lin,lin2,bf25,fb7} and quantum hall effect \cite{lin3,hc}. Besides, analogues of the Aharonov-Bohm effect for bound states \cite{ab,pesk} have been discussed for a neutral particles in several quantum systems \cite{fb7,ring4,ring7,barb3}.

In this work, we focus on a quantum system that consists of a neutral particle with an electric quadrupole moment that interacts with non-uniform electric fields. We analyse this system from a semiclassical point of view by using the the WKB (Wentzel, Kramers, Brillouin) approximation \cite{griff,landau,wkb4}. This approximation has been widely explored in the literature. For instance, it has been used to investigate anharmonic oscillators \cite{griff,wkb5,wkb8,wkb21}, inverse power-law potentials \cite{wkb6,wkb9,wkb15}, spin-orbit coupling \cite{wkb7}, polynomial potentials \cite{wkb11}, electron in a uniform magnetic field \cite{wkb14,wkb22}, $\mathcal{PT}$-symmetric quantum mechanics \cite{wkb16,wkb17,wkb18} and graphene \cite{wkb19,wkb20}. An interesting question about the use of WKB approximation was raised by Langer \cite{wkb} when a system has the spherical symmetry. Langer showed that it is needed to modify the centrifugal term in the radial equation, and how to deal with the singularity at the origin. This way of dealing with the centrifugal term and the singularity at the origin becomes known in the literature as the Langer modification or Langer transformation \cite{wkb2,wkb3,wkb4}. Based on the Langer transformation, systems with spherical symmetry have been investigated with the WKB approximation \cite{wkb,wkb4,wkb10,wkb12,wkb13}. Further, Berry and co-workers \cite{wkb2,wkb3} showed the treatment of the WKB approximation with the cylindrical symmetry. Therefore, in this work, we search for bound states solutions through the WKB approximation when the electric quadrupole moment of a neutral particle interacts with radial electric fields.

The structure of this paper is: in section II, we start with the quantum description of the electric quadrupole moment system. Then, we consider a long non-conductor cylinder that possesses non-uniform electric charge densities, and thus, make a semiclassical analysis of the interaction of the electric quadrupole moment and the radial electric fields produced by these non-uniform electric charge densities, in section III, we present our conclusions.

\section{Semiclassical analysis}

The electric quadrupole moment has brought attention in studies of atomic systems \cite{nucquad,nucquad2}, molecular systems \cite{quad1,quad,quad3,quad4,quad5}, nuclear structure \cite{quad2,quad6} and strongly magnetized Rydberg atoms \cite{prlquad}. With the focus on geometric quantum phases, Chen \cite{chen} obtained it from the interaction of a moving particle with electric quadrupole moment with a magnetic field. Then, based on the system proposed by Chen \cite{chen}, analogues of the scalar potential $V\left(r\right)\propto r^{-1}$ have been analysed in Refs. \cite{b3,b4}. An analogue of the Landau quantization has been studied in Ref. \cite{bf25}. On the other hand, the quantum description of the electric dipole moment system proposed in Ref. \cite{chen} allows to investigate the interaction between the electric quadrupole momentum of a neutral particle with electric fields. With this purpose, an analogue of the scalar Aharonov-Bohm effect has been obtained in Ref. \cite{b}. Moreover, based on the duality transformation or Heaviside transformations \cite{fur}, quantum effects associated with an analogue of the scalar potential $V\left(r\right)\propto r^{-1}$ that stems from the interaction of the electric quadrupole moment with an axial electric field have been investigated in Ref. \cite{b2}. In this section, our aim is to analyse the interaction of the electric quadrupole moment and radial electric fields through the WKB approximation \cite{wkb2,wkb3}.

Therefore, by considering a spinless neutral particle with an electric quadrupole moment that moves with velocity $v\ll c$ ($c$ is the velocity of light), the corresponding time-independent Schr\"odinger equation is given by \cite{b,b2,b3,b4}
\begin{eqnarray}
\mathcal{E}\psi=\frac{1}{2m}\left[\hat{p}+\frac{1}{c}(\vec{Q}\times\vec{B})\right]^2\,\psi-\vec{Q}\cdot\vec{E}\,\psi.
\label{1.1}
\end{eqnarray} 
Observe that the components of the vector $\vec{Q}$ are determined by $Q_{i}=\sum_{j}Q_{ij}\,\partial_{j}$, where $Q_{ij}$ corresponds to the electric quadrupole tensor. The tensor $Q_{ij}$ is symmetric and traceless. Moreover, the fields $\vec{B}$ and $\vec{E}$ are the magnetic and electric fields in the laboratory frame, respectively.

In the following, we analyse the interaction of the electric quadrupole moment of a neutral particle with electric fields produced by non-uniform electric charge densities inside a long non-conductor cylinder. This analysis is made through the WKB approximation based on the cylindrical symmetry \cite{wkb3,wkb4}.

\subsection{Analogue of the linear scalar potential}

Let us consider an electric field produced by a non-uniform (positive) electric charge density $\rho=\bar{\mu}\,r$ ($\bar{\mu}$ is a constant) inside of a long non-conductor cylinder. Thereby, the radial electric field is given by
\begin{eqnarray}
\vec{E}=\frac{1}{2}\mu\,r^{2}\,\hat{r},
\label{1.2}
\end{eqnarray}
where $\mu$ is a constant associated with the (positive) electric charge density. Furthermore, based on the properties of the tensor $Q_{ij}$, let us consider the electric quadrupole moment tensor to be defined by the following components:
\begin{eqnarray}
Q_{r\,r}=Q_{\varphi\,\varphi}=-Q;\,\,\,\,\,Q_{zz}=2Q,
\label{1.3}
\end{eqnarray}
where $Q$ is also a constant $\left(Q>0\right)$ and the remaining components are null. With Eqs. (\ref{1.2}) and (\ref{1.3}), we have that the interaction of the electric quadrupole moment of the neutral particle with the radial electric field yields an effective scalar potential in the Schr\"odinger equation (\ref{1.1}) given by
\begin{eqnarray}
V_{\mathrm{eff}}\left(r\right)=-\vec{Q}\cdot\vec{E}=Q\mu\,r.
\label{1.4}
\end{eqnarray}
Moreover, we have that this effective scalar potential plays the role of a linear scalar potential in the Schr\"odinger equation (\ref{1.1}).

With the cylindrical symmetry, hence, the time-independent Schr\"odinger equation for a neutral particle with an electric quadrupole moment (\ref{1.2}) that interacts with the radial electric field (\ref{1.3}) is written in the form:
\begin{eqnarray}
\mathcal{E}\psi=-\frac{\hbar^{2}}{2m}\,\left[\frac{\partial^{2}}{\partial r^{2}}+\frac{1}{r}\frac{\partial}{\partial r}+\frac{1}{r^{2}}\frac{\partial^{2}}{\partial\varphi^{2}}+\frac{\partial^{2}}{\partial z^{2}}\right]\psi+Q\,\mu\,r\,\psi.
\label{1.5}
\end{eqnarray}

Note that the operators $\hat{L}_{z}$ and $\hat{p}_{z}$ commute with the Hamiltonian operator given in the right-hand side of Eq. (\ref{1.5}), therefore, we can write the solution to the Schr\"odinger equation  (\ref{1.5}) in terms of the eigenvalues of these operators as follows:
\begin{eqnarray}
\psi\left(r,\,\varphi,\,z\right)=e^{il\varphi+ikz}\,R\left(r\right),
\label{1.6}
\end{eqnarray}
where $l=0,1,2,\ldots$ is the quantum number associated with the $z$-component of the angular momentum and $-\infty<\,k\,<\infty$ is the eigenvalue of the $z$-component of the linear momentum. Thereby, by substituting (\ref{1.6}) into Eq. (\ref{1.5}), we obtain the following radial equation:
\begin{eqnarray}
R''+\frac{1}{r}R'-\frac{l^{2}}{r^{2}}R-\frac{2m\,\mu\,Q}{\hbar^{2}}\,r\,R+\left[\frac{2m\mathcal{E}}{\hbar^{2}}\,-k^{2}\right]R=0.
\label{1.7}
\end{eqnarray}
From now on, let us write the radial wave function in the form:
\begin{eqnarray}
R\left(r\right)=\frac{1}{\sqrt{r}}\,u\left(r\right),
\label{1.8}
\end{eqnarray}
and then, the radial equation becomes
\begin{eqnarray}
u''-\frac{\left(l^{2}-1/4\right)}{r^{2}}u-\frac{2m\,\mu\,Q}{\hbar^{2}}\,r\,u+\left[\frac{2m\mathcal{E}}{\hbar^{2}}-k^{2}\right]u=0.
\label{1.9}
\end{eqnarray}

Based on Refs. \cite{wkb3,wkb4}, the WKB approximation becomes valid in the presence of the cylindrical symmetry when we replace the term $\left(l^{2}-1/4\right)$ with $l^{2}$ in the centrifugal term of the radial equation. Besides, let us simplify our discussion by taking $k=0$ from now on. In this way, we define
\begin{eqnarray}
q\left(r\right)=\sqrt{2m\left[\mathcal{E}-\mu\,Q\,r\right]-\frac{l^{2}\hbar^{2}}{r^{2}}},
\label{1.10}
\end{eqnarray}
and thus, the radial equation (\ref{1.9}) takes the form:
\begin{eqnarray}
u''+\frac{q^{2}\left(r\right)}{\hbar^{2}}\,u=0.
\label{1.11}
\end{eqnarray}

Furthermore, we must observe that the radial wave function is written in the WKB approach as \cite{wkb3,wkb4}:
\begin{eqnarray}
u\left(r\right)\cong\frac{2}{\sqrt{q\left(r\right)}}\,\cos\left(\frac{1}{\hbar}\,\int_{r_{1}}^{r}q\left(r'\right)\,dr'-\frac{\pi}{4}\right).
\label{1.12}
\end{eqnarray}

Hence, the Bohr-Sommerfeld quantization for the cylindrical symmetry is given by \cite{wkb3,wkb4}: 
\begin{eqnarray}
\frac{1}{\hbar}\,\int_{r_{1}}^{r_{2}}q\left(r\right)\,dr=\left(n-\frac{1}{2}\right)\pi,
\label{1.13}
\end{eqnarray}
where $n=1,2,3,\ldots$ is the quantum number associated with the radial modes, and $r_{1}$ and $r_{2}$ are the turning points.

An interesting point raised in Refs. \cite{wkb,wkb2,wkb3,wkb4} is the behaviour of the $s$ waves when we deal with the WKB approximation. The $s$ waves are defined by taking $l=0$. For $s$ waves, the wave function (\ref{1.12}) is well-behaved at $r_{1}=0$ as shown in Ref. \cite{wkb4}. Therefore, we have one turning point, which is determined when $\mathcal{E}=V\left(r_{2}\right)$. In this way, we have
\begin{eqnarray}
r_{2}=\frac{\mathcal{E}}{\mu\,Q}.
\label{1.14}
\end{eqnarray} 
With $r_{1}=0$ and $r_{2}$ given in Eq. (\ref{1.14}), then, the left-hand side of Eq. (\ref{1.13}) yields
\begin{eqnarray}
\frac{1}{\hbar}\,\int_{0}^{r_{2}}q\left(r\right)\,dr&=&\frac{1}{\hbar}\sqrt{2m\,\mu\,Q}\int_{0}^{r_{2}}\sqrt{r_{2}-r}\,dr\nonumber\\
[-2mm]\label{1.15}\\[-2mm]
&=&\frac{2}{3\,\hbar}\sqrt{2m\,\mu\,Q}\,\left(r_{2}\right)^{3/2}.\nonumber
\end{eqnarray}

By substituting Eq. (\ref{1.15}) into Eq. (\ref{1.13}), thus, we obtain that the allowed energies $\mathcal{E}_{n,\,l}$ associated with the $s$ waves: 
\begin{eqnarray}
\mathcal{E}_{n,\,0}=\mu\,Q\left[\frac{9\,\hbar^{2}\pi^{2}}{8m\,\mu\,Q}\left(n-\frac{1}{2}\right)^{2}\right]^{1/3}.
\label{1.16}
\end{eqnarray}

Hence, by using the WKB approximation to analyse the interaction of the electric quadrupole moment of a neutral particle (\ref{1.3}) with the radial electric field (\ref{1.2}), we have a discrete spectrum of energy that stems from this interaction. In this case, this interaction has gave rise to an analogue of the linear confining potential. However, the allowed energies (\ref{1.16}) are observed only for $s$ waves. For $l\neq0$, the turning point (\ref{1.14}) is modified and the result given in Eq. (\ref{1.15}) cannot be achieved. Moreover, the energy levels (\ref{1.16}) could not be obtained for other configuration of the electric quadrupole moment.

\subsection{Analogue of $r^{3}$-potential}

Let us consider another distribution of electric charges inside of a long non-conductor cylinder. The non-uniform electric charge density is given by $\rho=\bar{\nu}\,r^{3}$ ($\bar{\nu}$ is also a constant), thus, it produces inside the cylinder the following radial electric field:
\begin{eqnarray}
\vec{E}=\frac{1}{2}\nu\,r^{4}\,\hat{r},
\label{2.1}
\end{eqnarray}
where $\nu$ is also a constant associated with the (positive) electric charge density. Let us also consider the electric quadrupole moment of the neutral particle to be defined in Eq. (\ref{1.3}). Thereby, the effective scalar potential given in the Schr\"odinger equation (\ref{1.1}) becomes
\begin{eqnarray}
V_{\mathrm{eff}}\left(r\right)=-\vec{Q}\cdot\vec{E}=Q\nu\,r^{3}.
\label{2.2}
\end{eqnarray}
Since the radial coordinate is defined in the range $0\,<\,r\,<\,\infty$, then, we have that this effective scalar potential plays the role of a confining scalar potential in the Schr\"odinger equation (\ref{1.1}). In particular, the $r^{3}$-potential has been analysed in the literature as an anharmonic term with the purpose of obtaining the perturbative corrections of the spectrum of energy of the one-dimensional harmonic oscillator \cite{anharmonic,cohen,landau}. Therefore, with the effective scalar potential (\ref{2.2}), Eq. (\ref{1.1}) takes the form:
\begin{eqnarray}
\mathcal{E}\psi=-\frac{\hbar^{2}}{2m}\,\left[\frac{\partial^{2}}{\partial r^{2}}+\frac{1}{r}\frac{\partial}{\partial r}+\frac{1}{r^{2}}\frac{\partial^{2}}{\partial\varphi^{2}}+\frac{\partial^{2}}{\partial z^{2}}\right]\psi+Q\nu\,r^{3}\,\psi.
\label{2.2a}
\end{eqnarray}
By following the steps from Eq. (\ref{1.6}) to Eq. (\ref{1.10}), we can define:
\begin{eqnarray}
q\left(r\right)=\sqrt{2m\left[\mathcal{E}-\nu\,Q\,r^{3}\right]-\frac{l^{2}\hbar^{2}}{r^{2}}},
\label{2.3}
\end{eqnarray}
where we have taken $k=0$ and replaced the term $\left(l^{2}-1/4\right)$ with $l^{2}$ as in the previous section.

Let us also consider the $s$ waves, i.e., let us take $l=0$. As shown in Ref. \cite{wkb4}, for $s$ waves we have that $r_{1}=0$, thus, we have one turning point given when $\mathcal{E}=V\left(r_{2}\right)$. In this way, this turning point is
\begin{eqnarray}
r_{2}=\left(\frac{\mathcal{E}}{\nu\,Q}\right)^{1/3}.
\label{2.4}
\end{eqnarray}

In order to deal with the left-hand side of Eq. (\ref{1.13}), let us perform a change of variables given by $x=\frac{\nu\,Q}{\mathcal{E}}\,r^{3}$. Then, we have
\begin{eqnarray}
\frac{1}{\hbar}\,\int_{0}^{r_{2}}q\left(r\right)\,dr&=&\frac{\sqrt{2m\mathcal{E}}}{3\hbar}\left(\frac{\mathcal{E}}{Q\,\nu}\right)^{1/3}\,\int_{0}^{1}x^{-2/3}\,\sqrt{1-x}\,dx\nonumber\\
&=&\frac{\sqrt{2m\pi}}{6\hbar}\frac{1}{\left(Q\,\nu\right)^{1/3}}\,\frac{\Gamma\left(1/3\right)}{\Gamma\left(11/6\right)}\,\mathcal{E}^{5/6},
\label{2.5}
\end{eqnarray}
where $\Gamma\left(1/3\right)$ and $\Gamma\left(11/6\right)$ are the gamma functions \cite{arf}. Hence, by substituting Eq. (\ref{2.5}) into Eq. (\ref{1.13}), we obtain the allowed energies $\mathcal{E}_{n,\,l}$ associated with the $s$ waves:
\begin{eqnarray}
\mathcal{E}_{n,\,0}=\left[6\hbar\left(n-\frac{1}{2}\right)\sqrt{\frac{\pi}{2m}}\cdot\left(Q\,\nu\right)^{1/3}\cdot\frac{\Gamma\left(11/6\right)}{\Gamma\left(1/3\right)}\right]^{6/5}.
\label{2.6}
\end{eqnarray}

Hence, through the WKB approximation, we have that the $s$ waves of the neutral particle possesses a discrete spectrum of energy that stems from the effective scalar potential $V_{\mathrm{eff}}\left(r\right)\propto\,r^{3}$ that plays the role of a confining scalar potential. As we have seen, this effective scalar potential is produced by the interaction of the radial electric field (\ref{2.1}) and the electric quadrupole moment tensor defined in Eq. (\ref{1.3}). We should also observe that for other configurations of the electric quadrupole moment tensor $Q_{ij}$, the effective scalar potential (\ref{2.2}) could not be achieved and, as a consequence, the energy levels (\ref{2.6}) cannot be obtained. Besides, for $l\neq0$, the turning point (\ref{2.4}) is modified and the result given in Eq. (\ref{2.5}) cannot be achieved.

\subsection{Analogue of the attractive logarithmic potential}

Let us consider a non-uniform electric charge density inside of a long non-conductor cylinder, with an inner radius $r_{0}$, given by $\rho=E_{0}\,\ln\left(r/r_{0}\right)$, where $E_{0}$ is a constant ($E_{0}>0$). Then, this non-uniform electric charge density produces the radial electric field:
\begin{eqnarray}
\vec{E}=E_{0}\,r\left[\frac{1}{2}\,\ln\left(r/r_{0}\right)-\frac{1}{4}\right]\,\hat{r},
\label{3.1}
\end{eqnarray}

By considering the components of the electric quadrupole moment tensor to be defined in Eq. (\ref{1.3}), then, the effective scalar potential given in Eq. (\ref{1.1}) becomes
\begin{eqnarray}
V_{\mathrm{eff}}\left(r\right)=-\vec{Q}\cdot\vec{E}=\frac{Q\,E_{0}}{2}\left[\ln\left(r/r_{0}\right)-\frac{1}{2}\right].
\label{3.2}
\end{eqnarray}
Therefore, it plays the role of an attractive logarithmic potential \cite{wkb4,griff,log,log2} in the Schr\"odinger equation (\ref{1.1}). Thereby, Eq. (\ref{1.1}) becomes
\begin{eqnarray}
\mathcal{E}\psi=-\frac{\hbar^{2}}{2m}\,\left[\frac{\partial^{2}}{\partial r^{2}}+\frac{1}{r}\frac{\partial}{\partial r}+\frac{1}{r^{2}}\frac{\partial^{2}}{\partial\varphi^{2}}+\frac{\partial^{2}}{\partial z^{2}}\right]\psi+\frac{Q\,E_{0}}{2}\left[\ln\left(r/r_{0}\right)-\frac{1}{2}\right]\,\psi.
\label{3.2a}
\end{eqnarray}

Since we have the cylindrical symmetry in Eq. (\ref{3.2a}), thus, we can follow the steps from Eq. (\ref{1.6}) to Eq. (\ref{1.10}) and define:
\begin{eqnarray}
q\left(r\right)=\sqrt{2m\left[\bar{E}-\frac{Q\,E_{0}}{2}\ln\left(r/r_{0}\right)\right]-\frac{l^{2}\hbar^{2}}{r^{2}}},
\label{3.3}
\end{eqnarray}
where we have defined $\bar{E}=\mathcal{E}+\frac{q\,E_{0}}{4}$ and taken $k=0$.

As discussed in the previous section, for $s$ waves ($l=0$) we have that $r_{1}=0$. The turning point in the present case is
\begin{eqnarray}
r_{2}=r_{0}\,e^{2\bar{E}/Q\,E_{0}}.
\label{3.4}
\end{eqnarray} 

With the purpose of dealing with the left-hand side of Eq. (\ref{1.13}), let us perform a Langer-type transformation: $r=r_{0}\,e^{-x}$. Then, we have
\begin{eqnarray}
\frac{1}{\hbar}\,\int_{0}^{r_{2}}q\left(r\right)\,dr&=&\frac{r_{2}}{\hbar}\sqrt{m\,Q\,E_{0}}\,\int_{0}^{\infty}\sqrt{x}\,\,e^{-x}\,dx\nonumber\\
[-2mm]\label{3.5}\\[-2mm]
&=&\frac{r_{2}}{\hbar}\sqrt{m\,Q\,E_{0}}\,\frac{\sqrt{\pi}}{2}.\nonumber
\end{eqnarray}
Hence, by substituting Eq. (\ref{3.5}) into Eq. (\ref{1.13}), we can also obtain the energy levels associated with the $s$ waves:
\begin{eqnarray}
\mathcal{E}_{n,\,0}=\frac{Q\,E_{0}}{2}\,\ln\left(\frac{\hbar}{r_{0}}\sqrt{\frac{\pi}{Q\,E_{0}\,m}}\left[n-\frac{1}{2}\right]\right)-\frac{Q\,E_{0}}{4}.
\label{3.6}
\end{eqnarray}

Hence, by analysing the behaviour of the neutral particle subject to the effective attractive logarithmic potential through the WKB approximation, we have in Eq. (\ref{3.6}) the allowed energies associated with the $s$ waves. This effective attractive potential is produced by the the interaction of the radial electric field (\ref{3.1}) and the electric quadrupole moment tensor (\ref{1.3}). Note that effective scalar potential could not be achieved for an electric quadrupole tensor that differs from that defined in Eq. (\ref{1.3}). Furthermore, observe that for $l\neq0$ the turning point (\ref{3.4}) is different, thus, the results given in Eqs. (\ref{3.5}) and (\ref{3.6}) cannot be obtained.

\section{conclusions}

We have used the WKB approximation to analyse the interaction of the electric quadrupole moment of a neutral particle with radial electric fields produced by a non-uniform (positive) electric charge densities. Three different cases, where effective (attractive) scalar potentials stem from this interaction, have been proposed. In these cases, we have shown that bound states solutions associated with the $s$ waves can be achieved. Besides, we have obtained the allowed energies for each case. Observe that the non-uniform electric fields inside a long non-conductor cylinder are hard to achieve in the experimental context. In Ref. \cite{dop}, one can find discussions about this topic. On the other hand, this theoretical point of view of the interaction of the electric quadrupole moment with non-uniform electric fields can be in the interests of the studies of quantum effects on neutral particles systems. It has been inspired in Ref. \cite{er}, where a discussion about the possibility of achieving the quantum Hall effect for neutral particles with permanent magnetic dipole moment is made.

\acknowledgments

The author would like to thank CNPq for financial support.


\begin{thebibliography}{99}


\bibitem{ac} Y. Aharonov and A. Casher, Phys. Rev. Lett. {\bf53}, 319 (1984).

\bibitem{hmw} X.-G. He, B. h. J. McKellar, Phys. Rev. A {\bf47}, 3424 (1983).

\bibitem{hmw1} M. Wilkens, Phys. Rev. Lett. {\bf72}, 5 (1994).

\bibitem{whw} H. Wei, R. Han and X. Wei, Phys. Rev. Lett. {\bf75}, 2071 (1995)

\bibitem{whw2} H. Wei, X. Wei and R. Han, Phys. Rev. Lett. {\bf77}, 1657 (1996). 

\bibitem{chen} C.-C. Chen, Phys. Rev. A {\bf51}, 2611 (1995).

\bibitem{b} K. Bakke, Ann. Phys. (Berlin) {\bf524}, 338 (2012).

\bibitem{fb7} I. C. Fonseca and K. Bakke, Ann. Phys. (NY) {\bf363}, 253 (2015).

\bibitem{er} M. Ericsson and E. Sj\"oqvist, Phys. Rev. A {\bf65}, 013607 (2001).

\bibitem{lin} L. R. Ribeiro, C. Furtado and J. R. Nascimento, Phys. Lett. A {\bf348}, 135 (2006).

\bibitem{lin2} C. Furtado, J. R. Nascimento and L. R. Ribeiro, Phys. Lett. A {\bf358}, 336 (2006).

\bibitem{bf25} J. Lemos de Melo, K. Bakke and C. Furtado, Phys. Scr. {\bf84}, 045023 (2011).

\bibitem{lin3} L. R. Ribeiro, E. Passos, and C. Furtado, J. Phys. G: Nucl. Part. Phys. {\bf39}, 105004 (2012).

\bibitem{hc} M. F. Borunda {\it et al}, Phys. Rev. B {\bf78}, 245315 (2008).


\bibitem{ab} Y.  Aharonov  and D. Bohm,  Phys. Rev. {\bf115}, 485 (1959). 

\bibitem{pesk} M. Peshkin and A. Tonomura, \textit{The Aharonov-Bohm Effect} (Springer-Verlag, in: Lecture Notes in Physics, Vol. 340, Berlin, 1989).

\bibitem{ring4} A. V. Balatsky and B. L. Altshuler, Phys. Rev. Lett. {\bf70}, 1678 (1993).

\bibitem{ring7} S. Oh and C.-M. Ryu, Phys. Rev. B {\bf51}, 13441 (1995).


\bibitem{barb3} P. M. T. Barboza and K. Bakke, Ann. Phys. (NY) {\bf372}, 457 (2016). 

\bibitem{landau} L. D. Landau and E. M. Lifshitz, \textit{Quantum Mechanics, the nonrelativistic theory, 3rd Ed.} (Pergamon, Oxford, 1977).

\bibitem{griff} D. J. Griffiths, {\it Introduction to quantum mechanics, Second Edition}, (Prentice Hall, 2004).

\bibitem{wkb4} M. Brack and R. K. Bhaduri, {\it Semiclassical Physics} (Addison-Wesley Publishing Company, 1997).

\bibitem{wkb5} P. Gaudreau, R. M. Slevinsky and H. Safouhi, Ann. Phys. (NY) {\bf337}, 261 (2013).

\bibitem{wkb8} J. M. Cornwall and G. Tiktopoulos, Ann. Phys. (NY) {\bf228}, 365 (1993).

\bibitem{wkb21} R. Adhikari, R. Dutt and Y. P. Varshni, Phys. Lett. A {\bf131}, 217 (1998).

\bibitem{wkb6} F. M. Fern\'andez, Ann. Phys. (NY) {\bf379}, 83 (2017).

\bibitem{wkb9}  J. Trost and H. Friedrich, Phys. Lett. A {\bf228}, 127 (1997).

\bibitem{wkb15} H. Friedrich and J. Trost, Phys. Rev. A {\bf59}, 1683 (1999).

\bibitem{wkb7} H. Frisk and T. Guhr, Ann. Phys. (NY) {\bf221}, 229 (1993).

\bibitem{wkb11} A. Nanayakkara and I. Dasanayake, Phys. Lett. A {\bf294}, 158 (2002).

\bibitem{wkb14} A. Das, J. Frenkel, S. H. Pereira and J. C. Taylor, Phys. Rev. A {\bf70}, 053408 (2004).

\bibitem{wkb22} H. S. Yi, H. R. Lee and K. S. Sohn, Phys. Rev. A {\bf49}, 3277 (1994).

\bibitem{wkb16} C. M. Bender {\it et al}, J. Phys. A: Math. Gen. {\bf34}, L31 (2001).

\bibitem{wkb17} P. Dorey {\it et al}, J. Phys. A: Math. Gen. {\bf38}, 1305 (2005).

\bibitem{wkb18} C. M. Bender and H. F. Jones, Phys. Lett. A {\bf328}, 102 (2004).

\bibitem{wkb19} P. Delplace and G. Montambaux, Phys. Rev. B {\bf82}, 205412 (2010).

\bibitem{wkb20} Y. Zhang, Y. Barlas and K. Yang, Phys. Rev. B {\bf85}, 165423 (2012).

\bibitem{wkb} R. E. Langer, Phys. Rev. {\bf51}, 669 (1937).

\bibitem{wkb2} M. V. Berry and K. E. Mount, Rep. Prog. Phys. {\bf35}, 315 (1972).
 
\bibitem{wkb3} M. V. Berry and A. M. Ozorio de Almeida, J. Phys. A: Math., Nucl. Gen. {\bf6}, 1451 (1973).

\bibitem{wkb10} J. J. Morehead, J. Math. Phys. {\bf36}, 5431 (1995).

\bibitem{wkb12} Y. Ou, Z. Cao and Q. Shen, Phys. Lett. A {\bf318}, 36 (2013).

\bibitem{wkb13} J. Hainz and H. Grabert, Phys. Rev. A {\bf60}, 1698 (1999).

\bibitem{nucquad} R. Sternheimer, Phys. Rev. {\bf84}, 244 (1951).

\bibitem{nucquad2} R. Neugart and G. Neyens, {\it in The Euroschool Lectures on Physics with Exotic Beams, Vol. II}, Lect. Notes Phys. {\bf700}, Edited by J. Al-Khalili and E. Roeckl (Springer, Berlin Heidelberg 2006), p. 135-189.

\bibitem{quad1} A. D. Buckingham, R. L. Disch and D. A. Dunmur, J. Am. Chem. Soc. {\bf90}, 3104 (1968).

\bibitem{quad} J. H. Williams, Acc. Chem. Res. {\bf26}, 593 (1993).

\bibitem{quad3} A. D. Buckingham, J. Chem. Phys. {\bf30}, 1580 (1959).

\bibitem{quad4} G. Karl and J. D. Poll, J. Chem. Phys. {\bf46}, 2944 (1967).

\bibitem{quad5} R. LeSar and and D. R. Herschbach, J. Phys. Chem. {\bf87}, 5202 (1983).

\bibitem{quad2} G. Neyens, Rep. Prog. Phys. {\bf66}, 633 (2003).

\bibitem{quad6} G. Barone, R. Mastalerz, M. Reiher and R. Lindh, J. Phys. Chem. A {\bf112}, 1666 (2008).

\bibitem{prlquad} J.-H. Choi, J. R. Guest, A. P. Povilus, E. Hansist and G. Raithel, Phys. Rev Lett. {\bf95}, 243001 (2005).

\bibitem{b3} K. Bakke, Ann. Phys. {\bf341}, 86 (2014).

\bibitem{b4} K. Bakke, Eur. Phys. J. Plus {\bf130}, 129 (2015).

\bibitem{fur} C. Furtado and G. Duarte, Phys. Scr. {\bf71}, 7 (2005).

\bibitem{b2} K. Bakke, Int. J. Mod. Phys. A {\bf29}, 1450117 (2014).

\bibitem{anharmonic} J. E. Drummond, J. Phys. A: Math. Gen. {\bf14}, 1651 (1981).

\bibitem{cohen} C. Cohen-Tannoudji, B. Diu and F. Lalo\"e, {\it Quantum mechanics} (John Wiley, New York, 1977).

\bibitem{arf} G. B. Arfken and H. J. Weber, {\it Mathematical Methods for Physicists, sixth edition} (Elsevier Academic Press, New York, 2005).

\bibitem{log} C. Quigg and J. L. Rosner, Phys. Rep. {\bf56}, 167 (1979).

\bibitem{log2} M. Dineykhan, G. V. Efimov, G. Ganbold and S. N. Nedelko, {\it Oscillator Representation in Quantum Physics} (Springer, Berlin, 1995).

\bibitem{dop} W. E. Spear and P. G. Le Comber, Solid State Commun. {\bf17}, 1193 (1975). 



					


\end{thebibliography}
\end{document}